\documentclass{emulateapj}

\shorttitle{Rapid star formation in the presence of active galactic nuclei}
\shortauthors{Lintott \& Viti}

\begin{document}

\title{Rapid star formation in the presence of Active Galactic Nuclei}
\author{Chris Lintott\altaffilmark{1} and Serena Viti\altaffilmark{1}}
\altaffiltext{1}{Department of Physics and Astronomy, University College London, Gower Street, London, WC1E 6BT, UK}
\email{cjl@star.ucl.ac.uk}

\begin{abstract}

Recent observations reveal galaxies in the early Universe ($2<z<6.4$) with large reservoirs of molecular gas and extreme star formation rates. For a very large range of sources, a tight relationship exists between star formation rate and the luminosity of the HCN J=1-0 spectral line, but sources at redshifts of z$\sim$2 and beyond do not follow this trend. The deficit in HCN is conventionally explained by an excess of infrared (IR) radiation due to active galactic nuclei (AGN). We show in this letter not only that the presence of AGN cannot account for the excess of IR over molecular luminosity, but also that the observed abundance of HCN is in fact consistent with a population of stars forming from near-primordial gas.
\end{abstract}

\keywords{stars: formation --- galaxies: abundances --- galaxies: starburst ---astrochemistry}

\section{Introduction}

Recent work (Gao \& Solomon 2004, Wu et al. 2005) indicates that there exists a strong correlation between the luminosity of molecular lines \--- particularly the HCN J=1-0 transition\--- and the star formation rate derived from infrared luminosity. Impressively, these relations apply to individual star forming regions in the Milky Way and also to external systems across ten orders of magnitude in luminosity. However, the three highest redshift systems in the sample used by Gao \& Solomon (at $z\sim 2$) do not fit this trend, but have luminosities in HCN lower than would be expected given their star formation rate. A detection of the HCN J=5-4 line in APM 08279+5255 at a redshift z=3.91 has been reported with $L_{FIR}$/$L_{HCN}$ similar to that seen in other high-z sources (Wagg et al. 2005). Limits on HCN luminosities obtained for systems at much higher redshifts ($z \sim 6$) (Carilli et al. 2002) indicate that these systems also do not fit the low-z relation.

The conventional explanation for these discrepancies is to attribute the observed infrared excess to AGN activity. The $z\sim 2$ systems are, indeed, quasar hosts, presenting us with the problem of distinguishing IR radiation associated with star formation and that due to the AGN. Detailed modeling of the spectral energy distribution (SED) of one such source, the Cloverleaf quasar (Wei\ss~ et al. 2005), has shown that there exists a warm dust component which is believed to be associated with star-forming activity. In this letter we test the hypothesis that hidden AGN activity is responsible for the deviation of these high-z sources from the observed relationship between infrared and HCN luminosity. By using chemical models of high-mass star formation, we in fact find that AGN activity should be accompanied by an \emph{increase} in the relative abundance of HCN. Furthermore, we find that models of star formation from near-primordial gas, enriched only by the ejecta from population III stars, can account for the observed HCN abundance; hence we show that there is no `lack' of HCN at high redshift.

\section{The effect of AGN}

Our previous work (Lintott et al. 2005) used a time-dependent chemical model to simulate the formation of massive stars in order to constrain the expected emission from molecular species at high redshifts. Here, we make use of the same model to explore the effect of AGN on high redshift star forming systems. We use a model with initial atomic abundance ratios from calculations of the yield from zero-metallicity supernov\ae~(Chieffi \& Limongi 2004) and overall metallicity 1/100th solar. We then model the initial collapse of the protostellar clump, including chemical reactions both in the gas and on the surfaces of dust grains. Once a critical density is reached, we assume the high-mass protostar ignites, causing species trapped in the icy grain mantles to sublimate. The newly released molecules radiate strongly in the warm gas; the protostar has become a `hot core'. Although short-lived, hot cores are among the most luminous of all sources of sub-mm radiation in the Milky Way. Following previous work on low-mass star formation (Lepp \& Dalgarno 1996) we assume that the effect of AGN activity within the host galaxy will be an increase in ionization flux due to x-rays. 

Ionization is of crucial importance in driving the chemistry of star formation regions; the most significant gas phase reactions within dark clouds, particularly for oxygen and nitrogen bearing species, are ion-neutral reactions. We simulate the increase in x-ray ionization by enhancing the standard ($1.0\times 10^{17}\mathrm{s^{-1}}$) cosmic ray ionization rate by a factor, $\zeta$, of 2 to 1000.
The results of our modeling are shown in Figure 1.

\begin{figure}
\plotone{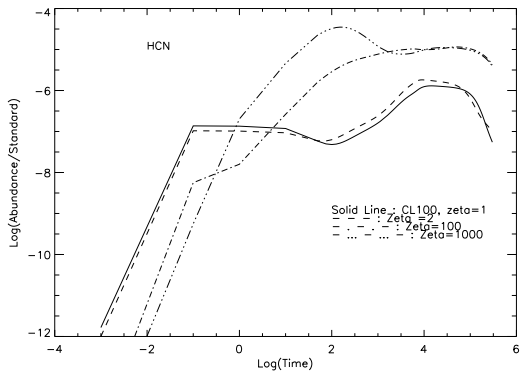}
\caption{Change in abundance for HCN relative to a `benchmark' model with only cosmic ray ionization and Milky Way metallicity and atomic abundances. The solid line represents a model with only cosmic ray ionization and abundances from Chieffi and Limongi (2004), standardized to 1/100th solar. Time since the ignition of the protostar and consequent sublimation of species from grain mantles shown on the x-axis.}
\end{figure}

The primary formation route for HCN in the benchmark model is via the reaction of $\mathrm{H_2}$ and CN. The latter is formed from atomic nitrogen and $\mathrm{C_2}$. UV photons formed as a by-product of the increased ionization produce enhanced abundances of atomic nitrogen, promoting the formation of CN and hence HCN. In addition, the increased ionization promotes the formation of HCN from dissociative recombination of $\mathrm{HCNH^+}$; this is the dominant formation route for HCN during the early stages of models with increased ionization. 
Note that reactions on the surface do not affect the abundance of HCN in our model
as there is no significant formation or destruction occuring on the grains. 
However, if freeze out is less efficient, more HCN would form in the gas phase 
during collapse.

A $\zeta$ of 100 results in an increase in relative abundance in HCN at all but very early times. Runs with $\zeta=1000$ produce an initially higher fractional abundance than those with $\zeta=100$, although the final abundance is unaffected. This result echoes that of Lepp and Dalgarno in low-mass star forming regions, although they find that, in these sources, the abundance of HCN (in common with all species in their model) declines at the high ionization rates explored by our model. In the model described here the increased density during collapse and the resulting freeze-out of molecules prevents dissociation.

Our models show clearly that an increase in HCN abundance is likely to accompany AGN activity. The magnitude of the increase (for relatively modest increases in ionization rate)  can be as large as two orders of magnitude. 
The far-IR excess seen in the most distant sources included by 
Gao \& Solomon is approximately an order of magnitude (a typical source 
with log($L_{\mathrm{HCN}}$)$\sim 9.5$ has $\log \left(L_{IR}\right)\sim 13.5$ 
while their relation would have predicted $\log \left(L_{IR}\right)\sim 12.5$).
Note that we do not model the physics of the AGN but only its probable chemical effects, and further research would be necessary to investigate the relationship between the increases in ionization and infrared luminosity. However, it seems likely that the consequent increase in HCN abundance from a AGN powerful enough to cause an order of magnitude increase in IR luminosity is large enough that the deviation from the observed trend cannot be explained by the presence of AGN. Similar enhancements of HCN in nearby AGN are observed (Kohno et al. 2001), and the use of HCN/HCO$^+$+ and HCN/CO ratios as diagnostics for AGN activity has been suggested. In addition, our preliminary results suggest that other species may also be worth investigating, in particular CS and $\mathrm{SO_2}$.

\section{IR excess or nitrogen under-abundance?}

If not AGN activity, what could be responsible for the deviation of molecular luminosities in high redshift systems from a relation which holds over a large range of luminosities? In a previous paper (Lintott et al. 2005) we suggested that the molecular gas in systems with rapid rates of star formation could be in the form of hot cores. Furthermore, gas enriched only by large zero-metallicity stars should be under-abundant in nitrogen compared to the solar neighbourhood. Hot cores formed from such gas should therefore have lower than expected abundances of nitrogen-bearing compounds. 

In this section, we investigate whether the molecular abundances derived in Section 2 are consistent with existing detections of HCN in high-redshift galaxies, and thus the hypothesis that deviations from the standard $L_{\mathrm{IR}}$/ $L_{\mathrm{HCN}}$ ratio are due not to an excess of infrared luminosity, but an under-abundance of HCN due to star formation from near-primordial gas.

We consider two sources observed using the Very Large Array (VLA), the Cloverleaf quasar (H1413, z=2.56) (Solomon et al. 2003) and J1409 (z=2.58, Carilli et al. 2002). Unfortunately, both of these sources have significant optical depths. However, we can still derive the column density as long as we assume a beam filling factor and excitation temperature (Nguyen-Q-Rieu et al. 1992). As the SED of the Cloverleaf has been studied in detail and a warm component with $T_d=50K$ associated with star formation detected, we will assume $T_{ex}=50K$ for both sources. We estimate the beam filling factor by assuming that the only source of HCN emission arises from hot cores with radii of 0.3pc. We have assumed such objects are larger than their Milky Way analogues in order to maintain an extinction large enough to harbour substantial reservoirs of molecular gas.

Our model predicts the contribution of each individual core to the column density, and, thus, given the number of cores in the source the overall column density. The observed luminosities for the Cloverleaf and J1409 in $\mathrm{K~kms^{-1}~pc^{2}}$ are $L_{HCN}=3.2\pm 0.5\times 10^{9}$ and $L_{HCN}=6.7\pm 2.2 \times 10^{9}$ respectively.

We find that our model is consistent with the observations in both cases if approximately $10^7$ hot cores are present. The column densities derived for the two sources are then $1.8\times 10^{17}\mathrm{cm^{-2}}$ (Cloverleaf) and $3.8\times 10^{16}\mathrm{cm^{-2}}$ (J1409). In Lintott et al. 2005 we estimated that the Milky Way contains $10^5$ hot cores and assert that the number of hot cores scales with star formation rate. As both of these sources have star formation rates (as determined by the infrared luminosities) greater than two orders of magnitude above the Milky Way rate, $10^7$ hot cores seems realistic. It should be noted that we have assumed that the observed molecular gas is entirely in the form of hot cores, and that these values should be considered as upper limits. It is clear from Figure 1 that the relative abundance of HCN in these models is at least a factor of $10^4$ below that expected in hot cores formed from gas with solar abundance ratios. Even if the number of hot cores in these rapidly star-forming systems is the same as in the Milky Way, hot cores with solar abundance ratios would result in a luminosity of HCN greatly in excess of the observed value. 

Thus there is no discrepancy between the star formation rate (derived from the IR luminosity) and the observed luminosity of HCN, if stars are forming from near-primordial gas at redshifts of z=2-4. 

\section{Conclusions}

We have argued that the observed differences between $L_{\mathrm{IR}}$/$L_{\mathrm{HCN}}$ at low and high redshifts cannot be due to the presence of AGN enhancement of the infrared luminosity. Such activity, we show, would increase the abundance of HCN in the observed systems. Detailed chemical models of such systems show promise in distinguishing between AGN enhanced and purely star forming activity through the ratio of HCN with species which are not enhanced by AGN. Examples of such species may include CS and $\mathrm{SO_2}$ (both of which are enhanced in galactic hot cores (Hatchell et al. 1998) and are expected to be favoured in nitrogen-deficient gas (Lintott et al. 2005). 

We explain the observations instead by hypothesizing that stars are forming from near-primordial material. Such material is believed to be deficient in nitrogen compared to the solar neighbourhood. 

Jiminez and Haiman (2006) discuss four observational constraints on the nature
of star formation at redshifts between 3 and 4. Specifically, they explain
significant UV emission in Ly-break galaxies, Ly-alpha emission isolated
from ionizing continuum radiation, a population of sources with strong
Ly-alpha emission and strong HeII lines by proposing a significant
fraction of star formation from primordial gas. These results imply that 
mixing of metals is inefficient within galaxies, and so solar abundance 
ratios may not have been reached at a redshift of z=2.

Such models present a challenge to existing models of chemical enrichment, which assume that the first population II stars form at much higher redshifts. Simulations (e.g. Abel et al. 2002) which suggest that population III stars will be massive assume that such stars will form at redshifts much greater than 4; it may be that the predictions of yields from such stars may need to be revised. 

\section*{Acknowledgments} CJL is supported by a PPARC studentship, and SV acknowledges individual financial support from a PPARC Advanced Fellowship.

\clearpage

\begin{table}
\begin{tabular}{|l|c|}
\hline
Reaction&Rate coefficient/$\mathrm{cm^3s^{-1}}$\\
\hline
H + HNC $\longrightarrow$ HCN + H&$6.2\times 10^{-10}\exp^{-\frac{12500}{T}}$\\
N + $\mathrm{CH_2}$ $\longrightarrow$ HCN + H& $7.89\times 10^{-11}\left(\frac{T}{300K}\right)^{0.17}$\\
$\mathrm{H_2}$ + CN $\longrightarrow$ HCN + H&$4.04\times 10^{-13}\left(\frac{T}{300K}\right)^{2.87}\exp^{-\frac{820}{T}}$\\
N + HCO $\longrightarrow$ HCN + O&$1.70\times10^{-10}$\\
$\mathrm{\textbf{HCNH}}^+ \mathrm{+ \textbf{e}}^- \longrightarrow \mathrm{\bf{HCN + H}}$&$9.0\times 10^{-8} \left(\frac{T}{300K}\right)^{-0.5}$\\
{\bf H +} $\mathrm{\textbf{H}}_2\mathrm{\textbf{CN}} \longrightarrow$ {\bf HCN +} $\mathrm{\textbf{H}}_2$&$1.0\times 10^{-10} \left(\frac{T}{300K}\right)^{0.5}$\\
\hline
$\mathrm{H_3^+}$ + HCN $\longrightarrow$ $\mathrm{HCNH^+}$ + $\mathrm{H_2}$&$8.1 \times 10^{-9}$\\
$\mathrm{H^+}$ + HCN $\longrightarrow$ $\mathrm{HCN^+}$ + H& $1.05\times 10^{-8} \left(\frac{T}{300K}\right)^{-0.13}$\\
$\mathrm{H_3O^+}$ + HCN $\longrightarrow$ $\mathrm{HCNH^+}$ + $\mathrm{H_2O}$&$4.0\times 10^{-9}$\\
{\bf HCN + Photon} + $\longrightarrow$ {\bf CN + H}& $1.3\times 10^{-9} \exp \left(-2.1A_v\right)$\\
\hline
\end{tabular}
\caption{Major formation and destruction reactions of HCN; all those which account for $10\%$ of the formation or destruction rate at any point in both benchmark and enhanced ionization models have been included. Those in bold are important only in models with enhanced ionization (although note the final reaction is important only in the very early stages). Rates are taken from the UMIST99 rate file. $T$ is the temperature and $A_v$ is the extinction.}
\end{table}


\begin{thebibliography}{99}

\bibitem{Abel} T. Abel, G.L. Bryan, M.L. Norman, {\it Science}, 295, 93 (2002)
\bibitem {Carilli} C.L. Carilli {\it et al., AJ}, 123, 1838 (2002)
\bibitem{CL04} A. Chieffi, M. Limongi, {\it ApJ}, 608, 405 (2004)
\bibitem{Gao} Y. Gao, P.M. Solomon, {\it ApJ}, 606, 271 (2004)
\bibitem{Jimenez} R. Jimenez, Z. Haiman, {\it Nature}, 440, 501 (2006)
\bibitem{Kohno} K. Kohno {\it et al.}, in `The central kpc of Starbursts and AGN: The La Palma Connection', J.H.Knapen, J.E. Beckman, I. Shlosman T.J.Mahoney Eds. (ASP Conf. Ser.), 249, 672 (2001)
\bibitem{Lepp} S. Lepp, A. Dalgarno, {\it A\&A}, 306, L21 (1996)
\bibitem{Lintott} C.J. Lintott, S. Viti, D.A. Williams, J.M.C. Rawlings, I. Ferreras, {\it MNRAS}, 260, 1527 (2005)
\bibitem{Nguyen-Q-Rieu} Nguyen-Q-Rieu, J.M. Jackson, C. Henkel, Truong-Bach, R. Mauersberger, ApJ, 399, 521 (1992)
\bibitem{Solomon} P. Solomon, P. Vanden Bout, C. Carilli, M. Guelin, {\it Nature}, 426, 636 (2003)
\bibitem{Wagg} J. Wagg, D.J. Wilner, R. Neri, D. Downes, T. Wiklind, {\it ApJ}, 634, L13 (2005)
\bibitem{Weiss} A. Wei\ss, D. Downes, C. Henkel, F. Walter, {\it A\&A}, 429, 25 (2005)
\bibitem{Wu} J. Wu {et al., ApJ}, 635, L173 (2005)

\end{thebibliography}
\end{document}